\documentclass{article}
\usepackage{spconf,amsmath,graphicx,url} % url追加 (高道)
\usepackage{booktabs}
\usepackage{subfigure}

\renewcommand{\Vec}[1]{\textrm{\boldmath $#1$}} % Vector

\newcommand{\pt}[1]{\left(#1\right)} % ()
\newcommand{\br}[1]{\left[#1\right]} % []
\title{Lifter training and sub-band modeling for computationally efficient and high-quality voice conversion using spectral differentials}
\makeatletter

\def\name#1{\gdef\@name{#1\\}}
\makeatother
\name{{\em Takaaki Saeki, Yuki Saito, Shinnosuke Takamichi, and Hiroshi Saruwatari}}
\address{
    Graduate School of Information Science and Technology, The University of Tokyo, Japan.
}

\begin{document}
\ninept
\maketitle
\setlength{\abovedisplayskip}{3pt} % 数式の上のマージン
\setlength{\belowdisplayskip}{3pt} % 数式の下のマージン
\renewcommand{\subfigcapskip}{1pt} %キャプションと図の間の間隔調整
\renewcommand{\subfigbottomskip}{1pt}
\setlength\floatsep{13pt} % 図と図の間
\setlength\textfloatsep{20pt} % 本文と図

% abstract
\begin{abstract} \vspace{-1mm}
In this paper, we propose computationally efficient and high-quality methods for statistical voice conversion (VC) with direct waveform modification based on spectral differentials. The conventional method with a minimum-phase filter achieves high-quality conversion but requires heavy computation in filtering. This is because the minimum phase using a fixed lifter of the Hilbert transform often results in a long-tap filter. One of our methods is a data-driven method for lifter training. Since this method takes filter truncation into account in training, it can shorten the tap length of the filter while preserving conversion accuracy. Our other method is sub-band processing for extending the conventional method from narrow-band (16~kHz) to full-band (48~kHz) VC, which can convert a full-band waveform with higher converted-speech quality. Experimental results indicate that 1) the proposed lifter-training method for narrow-band VC can shorten the tap length to $1/16$ without degrading the converted-speech quality and 2) the proposed sub-band-processing method for full-band VC can improve the converted-speech quality than the conventional method.
\end{abstract}

% keyword
\vspace{-1mm}

% introduction
\vspace{-1mm}
\section{Introduction}\label{intro}
\vspace{-1mm}

Voice conversion (VC) is a method for converting the characteristics of a source speech into those of a target speech, while keeping the linguistic information unchanged~\cite{abe88}. VC has the potential to achieve speech communication beyond the physical constraints of the human vocal organs~\cite{toda04_augumented}. The most common VC method is statistical VC, which constructs an acoustic model that converts speech features of a source speaker into those of a target speaker. Deep neural network (DNN)-based VC~\cite{desai09nnvc, sun15} has been widely studied, and many models for achieving higher-converted-speech quality have been proposed. From a practical point of view, real-time VC methods based on a Gaussian mixture model~\cite{toda12} and DNN~\cite{arakawa19} have also been studied. They achieve online high-quality conversion of narrow-band (16~kHz) speech using a single CPU on a laptop PC. However, their computational cost is still high, and we need to reduce this cost towards portable (e.g., VC using a low-power CPU on a smart phone) or full-band (48~kHz) VC.

VC consists of three steps: feature analysis, feature conversion, and waveform synthesis. We particularly focus on the last step and use a spectral-differential VC~\cite{kobayashi18} that performs VC in the waveform-domain by applying a spectral differential filter to the source speech waveform. This 1) achieves high-quality conversion by avoiding vocoder errors and 2) incurs less computational cost than neural vocoders~\cite{tamamori17wavenetvocoder, kalchbrenner18wavernn, Wang2018NeuralSW} that use large DNNs and require sample-by-sample heavy computation. The spectral-differential VC originally used a mel-log spectrum approximation (MLSA) filter~\cite{imai83mlsa} to filter a source speech, but Suda et al. found that using a minimum-phase filter achieved higher converted-speech quality than using the MLSA filter~\cite{suda18}. In the case of the minimum-phase filter, an acoustic model (e.g., DNN) outputs a real cepstrum of the converted speech, and the Hilbert transform using a lifter with fixed parameters determines the phases of the filter from the real cepstrum. These processes are suitable for our aim because their computational costs (i.e., filter design) are very small. However, since the minimum-phase filter is not guaranteed to have a short tap length (i.e., the number of samples of the filter), it increases the computational cost of filtering. The practical way to reduce the cost is to truncate the filter in the time domain~\cite{sunohara17}, e.g., using the first half taps instead of full taps. However, such filter truncation degrades converted-speech quality. 

\begin{figure}[t]
  \centering
  \includegraphics[width=0.80\linewidth, clip]{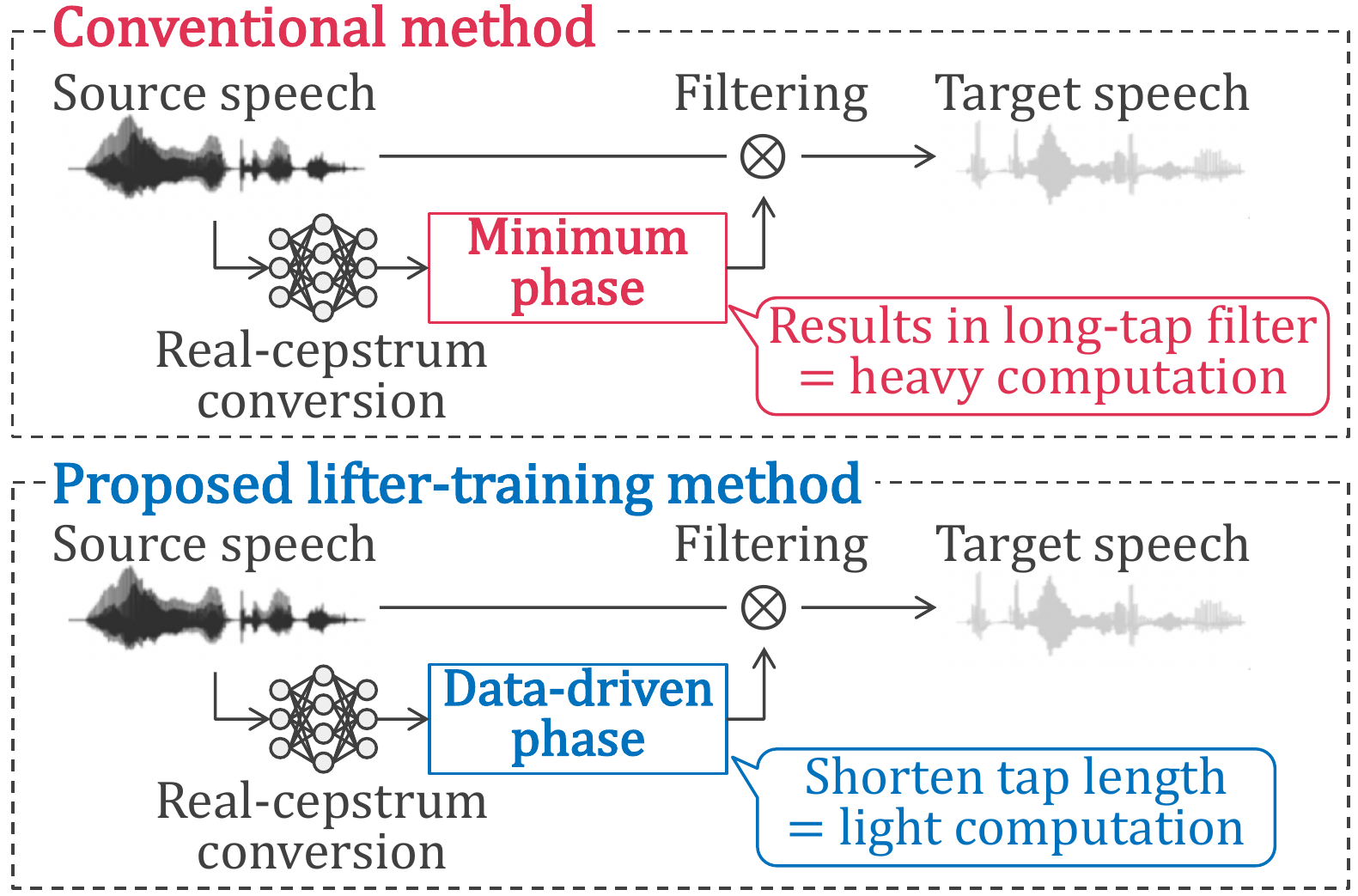}
  \vspace{-8pt}
  \caption{Comparison of conventional and proposed lifter-training methods}
  \vspace{-10pt}
  \label{fig:concept}
\end{figure}

Therefore, we propose a lifter-training method for reducing computational cost without degrading converted-speech quality. Our method jointly trains not only a DNN-based acoustic model but also a lifter with trainable parameters. Since parameters of the DNNs and the lifter are optimized to maximize conversion accuracy with the consideration of a truncated (i.e., short-tap) filter, our method can reduce the computational cost while preserving conversion accuracy. The main difference between our method and a conventional one using a minimum-phase filter is with the lifter to determine the phase of the filter, as shown in Fig.~\ref{fig:concept}. Whereas the lifter of the minimum-phase filter is \textit{fixed}, that of our method is \textit{trained} from speech data to determine the phases of a truncated filter. Furthermore, in this paper we extend the conventional method from narrow-band (16~kHz) to full-band (48~kHz) VC. Since fluctuations in wider-band voices are difficult to model with statistical models, the quality of the converted speech is relatively bad. Thus, we also propose a sub-band-processing method for improving the converted-speech quality. This method is designed to statistically convert the lower frequency band and preserve the higher frequency band. We conducted objective and subjective evaluations to investigate the effectiveness of the two proposed methods. Experimental results indicate that 1) the proposed lifter-training method for narrow-band VC can shorten the tap length to $1/16$ without degrading converted-speech quality and 2) the proposed sub-band-processing method for full-band VC can improve the converted-speech quality.

\vspace{-1mm}
\section{Conventional spectral-differential VC with minimum-phase filter}
\vspace{-1mm}
This section describes the training and conversion processes of the conventional spectral-differential VC with a minimum-phase filter.

\vspace{-1mm}
\subsection{Training process}\label{conv_train}
\vspace{-1mm}
Let $\Vec{F}^{(\mathrm{X})} = [{\Vec{F}^{(\mathrm{X})}_{1}}^{\top},...,{\Vec{F}^{(\mathrm{X})}_{t}}^{\top},...,{\Vec{F}^{(\mathrm{X})}_{T}}^{\top}]^{\top}$ be a complex frequency spectrum sequence obtained by applying the short-time Fourier transform (STFT) to an input speech waveform, where $t$ represents the frame index and $T$ is the total number of frames.
For the simplicity, we now focus on the frame $t$. A low-order real cepstrum $\Vec{C}_{t}^{(\mathrm{X})}$ can be extracted from $\Vec{F}_{t}^{(\mathrm{X})}$~\cite{fukada92melcep}. The DNNs then estimate a real cepstrum of differential filter $\Vec{C}_{t}^{(\mathrm{D})}$ from $\Vec{C}_{t}^{(\mathrm{X})}$. The loss function for $t$ is calculated as $L_{t} = (\Vec{C}_{t}^{(\mathrm{Y})} - \hat{\Vec{C}}_{t}^{(\mathrm{Y})})^{\top} 
     (\Vec{C}_{t}^{(\mathrm{Y})} - \hat{\Vec{C}}_{t}^{(\mathrm{Y})})$, where $\hat{\Vec{C}}_{t}^{(\mathrm{Y})}$ is a real cepstrum of converted speech given as $\hat{\Vec{C}}_{t}^{(\mathrm{Y})} =  \Vec{C}_{t}^{(\mathrm{X})} + \Vec{C}_{t}^{(\mathrm{D})}$, and $\Vec{C}_{t}^{(\mathrm{Y})}$ is a real cepstrum of the target speech. The DNNs are trained to minimize the loss function for all time frames represented as follows:
\begin{equation}
    \label{L_conv}
    L = \frac{1}{T} \sum_{t=1}^{T} L_{t}.
\end{equation}

\vspace{-1mm}
\subsection{Conversion process}
\vspace{-1mm}
$\Vec{C}_{t}^{(\mathrm{D})}$ is estimated with the DNNs. After the high-order components of the cepstrum are padded with zeros, $\Vec{C}_{t}^{(\mathrm{D})}$ is multiplied by a time-independent lifter $\Vec{u}_{\rm{min}}$ for a minimum-phase filter. The complex frequency spectrum of differential filter $\Vec{F}_{t}^{(\mathrm{D})}$ can be obtained by taking the inverse discrete Fourier transform (IDFT) of the liftered cepstrum.
The lifter $\Vec{u}_{\rm{min}}$ is represented as follows~\cite{pei06}:
\begin{equation}
\label{minphase}
  \Vec{u}_{\rm{min}}(n) = \begin{cases}
    1 & \pt{n=0, n = N/2} \\
    2 & \pt{0 < n < N/2}, \\
    0 & \pt{n > N/2}
  \end{cases}
\end{equation}
where $N$ is the number of frequency bins of the DFT. A differential filter in the time domain $\Vec{f}_{t}^{(\mathrm{D})}$ is obtained by applying the IDFT to $\Vec{F}_{t}^{(\mathrm{D})}$. The tap length of $\Vec{f}_{t}^{(\mathrm{D})}$ is equal to $N$. To reduce the computational cost of convolution operation, we can truncate $\Vec{f}_{t}^{(\mathrm{D})}$ with a fixed tap length $l$ ($l < N$). We define the $l$-tap truncated filter as $\Vec{f}_{t}^{(l)}$. Although filter truncation can efficiently reduce the computational cost, $\Vec{f}_{t}^{(l)}$ degrades converted-speech quality.

\vspace{-1mm}
\section{Proposed methods}
\vspace{-1mm}

We present the proposed methods for lifter training with filter truncation for computational-cost reduction and sub-band processing for improving the full-band converted-speech quality.

\vspace{-1mm}
\subsection{Lifter training with filter truncation}
\begin{figure}[t]
  \centering
  \includegraphics[width=0.9\linewidth, clip]{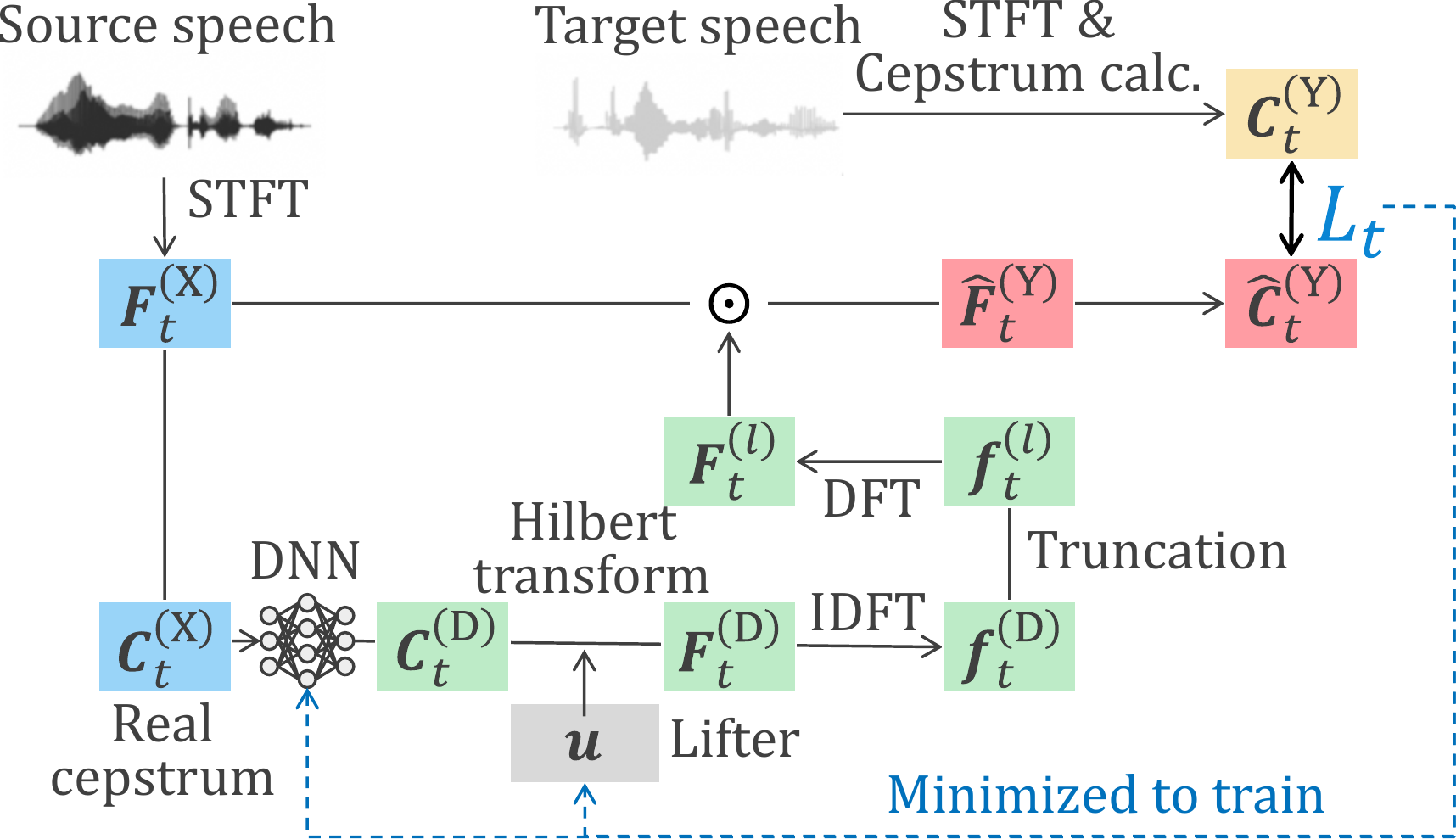}
  \vspace{-8pt}
  \caption{Procedure of proposed lifter-training method}
  \vspace{-10pt}
  \label{fig:proposed_training}
\end{figure}

\vspace{-1mm}
Our lifter-training method trains not only DNNs but also a lifter to avoid converted-speech quality degradation caused by filter truncation. Let $\Vec{u} = [u_{1},...,u_{c}]^{\top}$ be a time-independent trainable lifter, where $c$ is the dimension of the real cepstrum. The filter-truncation process with $l$ is integrated into training as shown in Fig.~\ref{fig:proposed_training}. 

As we described in Section~\ref{conv_train}, the DNNs estimate $\Vec{C}_{t}^{(\mathrm{D})}$ from $\Vec{C}_{t}^{(\mathrm{X})}$. Then $\Vec{C}_{t}^{(\mathrm{D})}$ is multiplied by the trainable lifter $\Vec{u}$, and the complex frequency spectrum of the differential filter $\Vec{F}_{t}^{(\mathrm{D})}$ is obtained from the IDFT of $\Vec{C}_{t}^{(\mathrm{D})}$ and exponential calculation. The differential filter in the time domain $\Vec{f}_{t}^{(\mathrm{D})}$ is obtained by applying the IDFT to $\Vec{F}_{t}^{(\mathrm{D})}$. $\Vec{f}_{t}^{(\mathrm{D})}$ is truncated to $\Vec{f}_{t}^{(l)}$ by applying a window function $\Vec{w}$ given as Eq.~(\ref{truncation}):
\begin{eqnarray}
    \label{truncation}
    \Vec{f}_{t}^{(l)} &=& \Vec{f}_{t}^{(\mathrm{D})} \cdot \Vec{w}, \\
    \Vec{w}           &=& \br{\overset{0\mathrm{th}}{1}, \cdots, \overset{(l-1)\mathrm{th}}{1}, \overset{l\mathrm{th}}{0}, \cdots, \overset{(N-1)\mathrm{th}}{0}}^\top.
\end{eqnarray}
% \begin{equation}
% \label{truncation}
%     \Vec{f}_{t}^{(l)} = \Vec{f}_{t}^{(\mathrm{D})} \cdot \Vec{w} 
% \end{equation}
By using the DFT again, a complex spectrum of the $l$-tap truncated differential filter $\Vec{F}_{t}^{(l)}$ can be obtained. A complex spectrum of converted speech $\hat{\Vec{F}}_{t}^{(\mathrm{Y})}$ is obtained by multiplying $\Vec{F}_{t}^{(\mathrm{X})}$ by $\Vec{F}_{t}^{(l)}$, and the real cepstrum of converted speech $\hat{\Vec{C}}_{t}^{(\mathrm{Y})}$ is extracted from $\hat{\Vec{F}}_{t}^{(\mathrm{Y})}$. The parameters of the DNNs and the lifter are jointly trained to minimize the same loss function as Eq.~(\ref{L_conv}).
Since all processes of this method are differentiable, the training can be done by back-propagation~\cite{backprop}.

\vspace{-1mm}
\subsection{Conversion process}
\vspace{-1mm}
In the conversion process, the trained DNNs and lifter estimate $\Vec{F}_{t}^{(\mathrm{D})}$. $\Vec{f}_{t}^{(\mathrm{D})}$ is obtained by applying the IDFT to $\Vec{F}_{t}^{(\mathrm{D})}$, and $\Vec{f}_{t}^{(l)}$ is obtained by truncating with $l$. We can obtain the converted speech waveform by applying $\Vec{f}_{t}^{(l)}$ to the source speech waveform.

\subsection{Sub-band processing for full-band VC}
We now describe our sub-band processing method for full-band extension of the conventional method. When the bandwidth of spectral-differential VC is extended to 48~kHz, conversion cannot be performed well due to the large fluctuations in the wider-band components. To avoid errors in the high-frequency components, our method involves sub-band processing for filtering separately for each frequency band. We apply the differential filter in a frequency region lower than 8~kHz, and do not apply it to a frequency region higher than 8~kHz. In practice, conversion of only the low-frequency band is possible by 1) subtracting 1.0 from the absolute value of $\Vec{F}_{t}^{(\mathrm{D})}$, 2) applying the sigmoid function that transitions 1 to 0 at around 8~kHz, and 3) adding 1.0 again.

\vspace{-1mm}
\subsection{Discussion}
\begin{figure}[t]
  \centering
  \includegraphics[width=0.85\linewidth, clip]{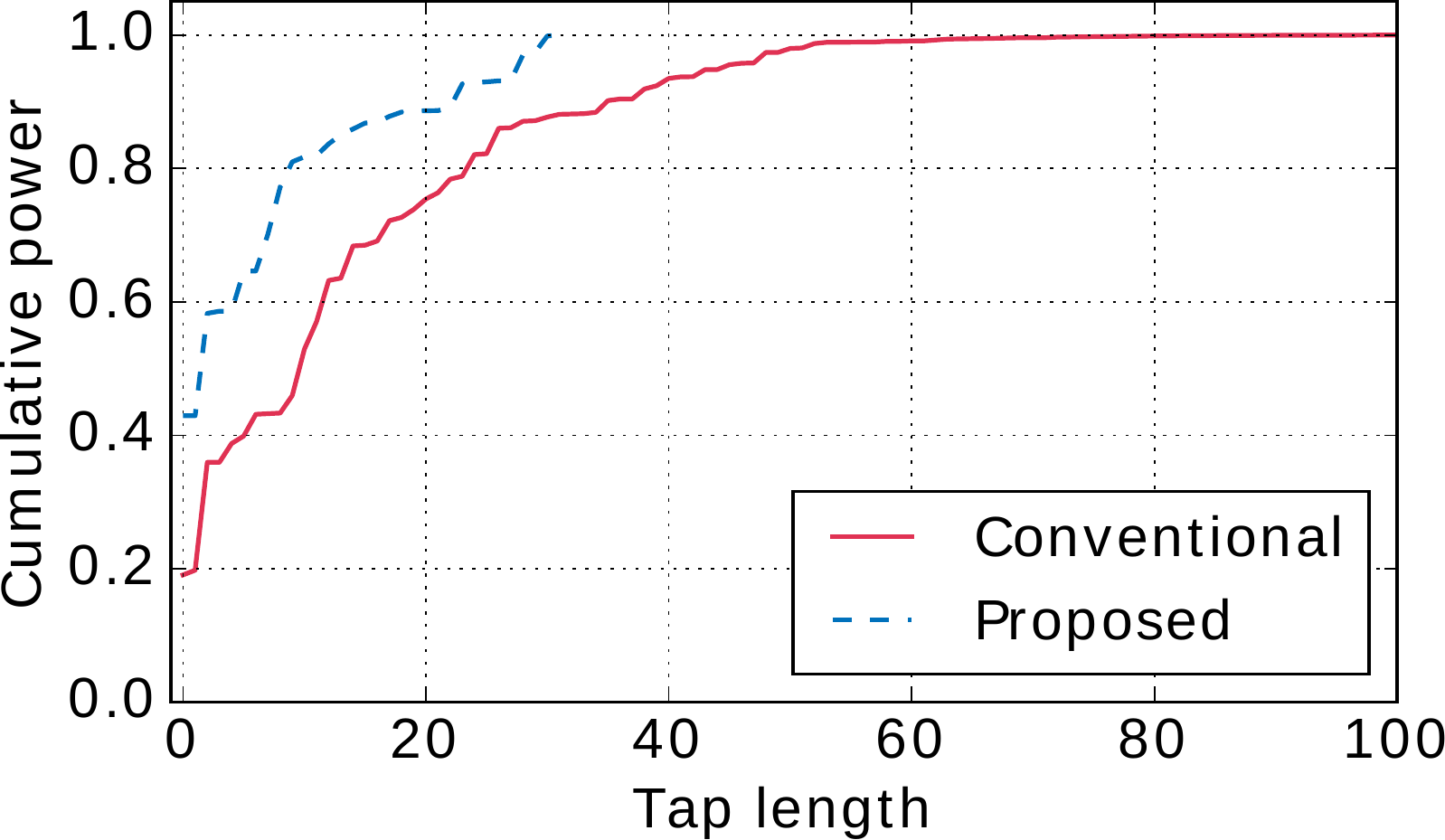}
  \vspace{-8pt}
  \caption{Cumulative power distributions of the differential filter}
  \vspace{-8pt}
  \label{fig:cumpow}
  %\vspace{10pt}
\end{figure}

\begin{figure}[t]
  \centering
  \includegraphics[width=0.85\linewidth, clip]{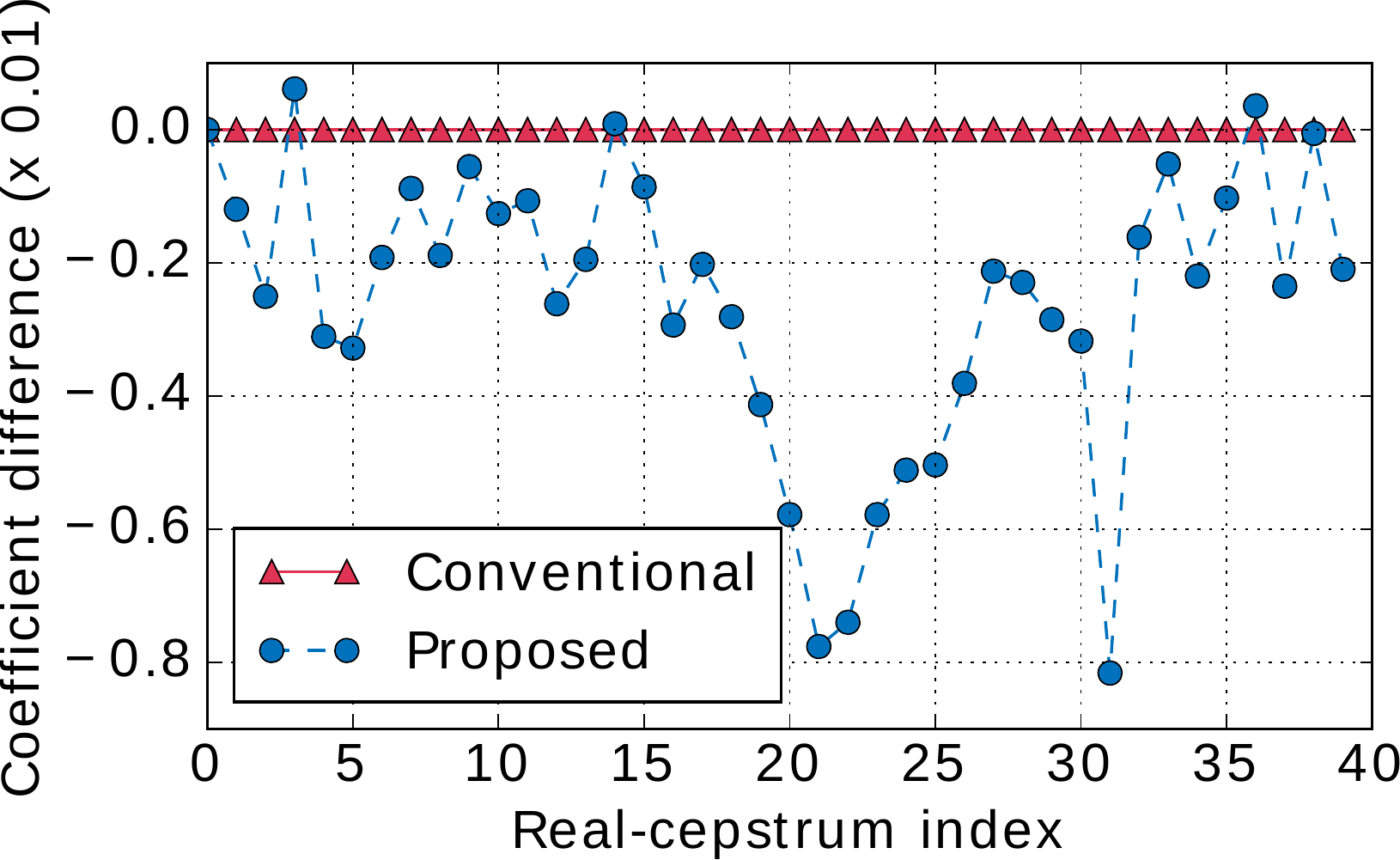}
  \vspace{-8pt}
  \caption{Difference between lifter trained with the proposed lifter-trained method ($l=64$) and that for minimum phasing}
  \vspace{-8pt}
  \label{fig:lifter_diff}
\end{figure}

\vspace{-1mm}
With the conventional method, the cepstrum is multiplied by the lifter coefficient to determine the shape of the filter so that the phase is minimized. Although the shape of the differential filter changes due to truncation, it is transformed to compensate for the effect of the truncation by applying the Hilbert transform using the lifter trained with the proposed lifter-training method. As a result, our lifter-training method can reduce the calculation amount while suppressing converted-speech quality degradation caused by the filter truncation. Figure~\ref{fig:cumpow} shows the cumulative power distribution of the differential filter with the conventional method ($l=512$) and  proposed lifter-training method ($l=32$). The values on the vertical axis are normalized with the cumulative total. The power is concentrated in the region of tap length 0 to 100. Figure~\ref{fig:lifter_diff} also shows the difference between the lifter trained with the proposed method ($l=64$) and that for minimum phasing. 

As explained in Section~\ref{intro}, liftering-based phase estimation requires only small computation. Since our lifter-training method adopts the same estimation as the conventional method, there is no increase in computational cost of phase estimation. 

We applied our lifter-training method to VC, i.e., speaker conversion. We expect that this method can be applied to other tasks processed by filtering, e.g., source separation and speech enhancement.

We also discuss our sub-band processing method. Since the characteristics of a speech waveform vary significantly from band to band, it is effective to process the waveform separately for each band. In sub-band WaveNet~\cite{okamoto17}, the speech waveform is divided into several bands and down-sampled, and the waveform in each band is processed separately. In our proposed sub-band processing method using full-band speech, the waveform of each frequency band is not down-sampled. In implementing real-time conversion, the computational cost of filtering can be reduced to $1/3$ by dividing the frequency domain into three bands and down-sampling the waveform of each band.

 \vspace{-1mm}
\section{Experimental evaluations}
\vspace{-1mm}

\subsection{experimental condition}
\vspace{-1mm}
We built two intra-gender VC: for female-to-female (f2f) and male-to-male (m2m) conversion. The source and target speakers in female-to-female conversion were stored in the JSUT corpus~\cite{jsut} and Voice Actress Corpus~\cite{vsdataset}, respectively. Those in male-to-male conversion were stored in the JVS corpus~\cite{jvs}. We used 100 utterances (approx. 12 min.) of each speaker, and the numbers of utterances for training, validation, and test data were 80, 10, 10, respectively.

We used narrow-band speech (16~kHz) and full-band speech (48~kHz) for the evaluation. In the narrow-band case, the window length was 25~ms, frame shift was 5~ms, the fast Fourier transform (FFT) length was 512~samples, and number of dimensions of the cepstrum was 40 (0th-through-39th). In the full-band case, the window length and frame shift were the same as those in the narrow-band case, but the FFT length was 2048 samples, and number of dimensions of the cepstrum was 120 (0th-through-119th). For pre-processing, the silent intervals of training and validation data were removed, and the lengths of the source and target speech were aligned by dynamic time warping.

The DNN architecture of the acoustic model was multi-layer perceptron consisting of two hidden layers. The numbers of each hidden unit were 280 and 100 in the narrow-band case, and 840 and 300 in the full-band case. The DNNs consisted of a gated linear unit~\cite{dauphin16gatedlinear} including the sigmoid activation layer and tanh activation layer, and batch normalization~\cite{ioffe2015batch} was carried out before applying each activation function. Adam~\cite{kingma14adam} was used as the optimization method. During training, the cepstrum of the source and target speech was normalized to have zero mean and unit variance. The batch size and number of epochs were set to 1,000 and 100, respectively. The model parameters of the DNNs used with the proposed lifter-training method were initialized with the conventional method. The initial value of the lifter coefficient was set to that of the lifter for minimum phasing. In the narrow-band case, the learning rates for the conventional and proposed lifter-training methods were set to 0.0005 and 0.00001, respectively. In the full-band case, the learning rates for the conventional and proposed lifter-training methods were set to 0.0001 and 0.000005, respectively. 

The proposed lifter-training method was evaluated using both narrow-band (16~kHz) and full-band (48~kHz) speech. The truncated tap length $l$ for the narrow-band case was 128, 64, 48, and 32, and that for the full-band case was 224 and 192. On the other hand, the proposed sub-band processing method was evaluated using only the full-band speech.

\vspace{-1mm}
\subsection{Objective evaluation}
\begin{figure}[t]
  \centering
  \includegraphics[width=0.85\linewidth, clip]{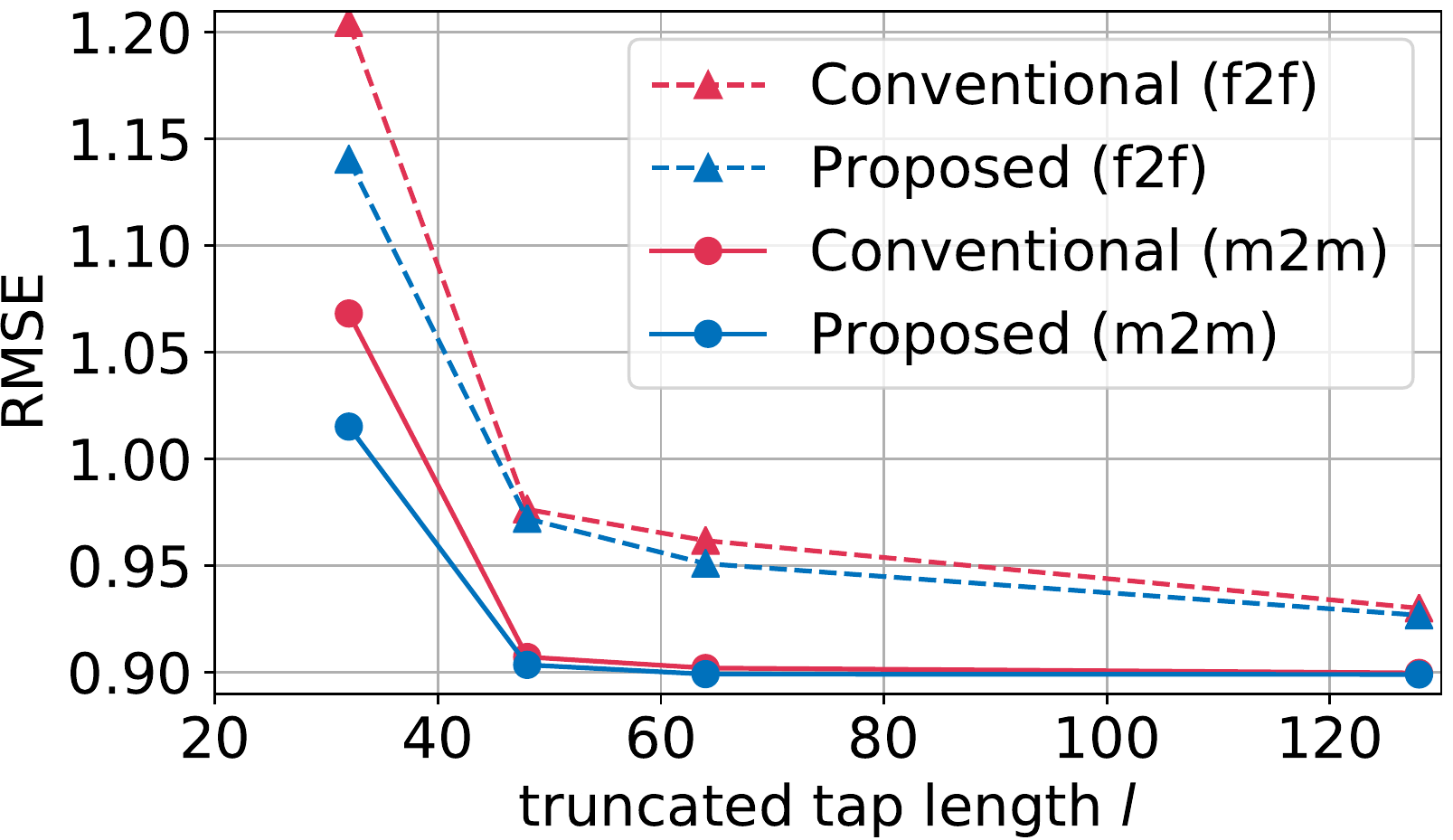} 
  \vspace{-8pt}
  \caption{RMSEs at each $l$ in narrow-band case \textbf{(16~kHz)}}
  \vspace{-8pt}
  \label{fig:test_loss}
\end{figure}

\vspace{-1mm}
We compared root mean squared error (RMSE) with the proposed lifter-training and conventional methods when changing $l$. The RMSE is obtained by taking the squared root of Eq.~(\ref{L_conv}). Figure~\ref{fig:test_loss} shows a plot of the RMSEs in male-to-male and female-to-female VC using narrow-band speech (16~kHz). The proposed lifter-training method achieved higher-precision conversion than the conventional method for all $l$. The differences in the RMSEs between the proposed and conventional methods also tended to become more significant when $l$ is smaller. This result indicates that the proposed lifter-training method can reduce the effect of filter truncation.

\setlength{\abovedisplayskip}{3pt} % 数式の上のマージン
\setlength{\belowdisplayskip}{3pt} % 数式の下のマージン
\renewcommand{\subfigcapskip}{1pt} %キャプションと図の間の間隔調整
\renewcommand{\subfigbottomskip}{1pt}
\setlength\floatsep{8pt} % 図と図の間
\setlength\textfloatsep{8pt} % 本文と図

\begin{table}[tb]\label{tab:eval-16kHz}
    \centering
    \caption{Preference scores with proposed lifter-training and conventional methods in narrow-band case \textbf{(16~kHz)}}
    \vspace{1mm}
    \subtable[\textbf{Speaker similarity}]{
        \footnotesize
        \begin{tabular}{r|cl|l}
            \hline \hline
            Proposed & Score & $p$-value & Conventional \tabularnewline
            \hline
            $l = 32$ (m2m) & \textbf{0.587} vs. 0.413 & $1.3 \times 10^{-5}$ & $l = 32$ (m2m) \tabularnewline
            $l = 32$ (m2m) &    0.463       vs. 0.537 & $7.3 \times 10^{-2}$ & $l = 512$ (m2m) \tabularnewline
            $l = 32$ (f2f) & \textbf{0.642} vs. 0.358 & $ < 10^{-10}$ & $l = 32$ (f2f) \tabularnewline
            $l = 32$ (f2f) & \textbf{0.543} vs. 0.457 & $3.4 \times 10^{-2}$ & $l = 512$ (f2f) \tabularnewline
            $l = 48$ (m2m) & 0.533 vs. 0.467 & $1.0 \times 10^{-1}$ & $l = 48$ (m2m) \tabularnewline
            $l = 48$ (m2m) & \textbf{0.550} vs. 0.450 & $1.4 \times 10^{-2}$ & $l = 512$ (m2m) \tabularnewline
            $l = 48$ (f2f) & \textbf{0.613} vs. 0.387 & $1.3 \times 10^{-8}$ & $l = 48$ (f2f) \tabularnewline
            $l = 48$ (f2f) & \textbf{0.548} vs. 0.452 & $2.0 \times 10^{-2}$ & $l = 512$ (f2f) \tabularnewline
            \hline \hline
        \end{tabular}
    }
    \subtable[\textbf{Speech quality}]{
        \vspace{-3mm}
        \footnotesize
        \begin{tabular}{r|cl|l}
            \hline \hline
            Proposed & Score & $p$-value & Conventional \tabularnewline
            \hline
            $l = 32$ (m2m) &    \textbf{0.687}       vs. 0.313 & $ < 10^{-10}$ & $l = 32$ (m2m) \tabularnewline
            $l = 32$ (m2m) &    0.529       vs. 0.471 & $2.3 \times 10^{-1}$ & $l = 512$ (m2m) \tabularnewline
            $l = 32$ (f2f) & \textbf{0.807} vs. 0.193 & $ < 10^{-10}$ & $l = 32$ (f2f) \tabularnewline
            $l = 32$ (f2f) & \textbf{0.742} vs. 0.258 & $ < 10^{-10}$ & $l = 512$ (f2f) \tabularnewline
            $l = 48$ (m2m) &    \textbf{0.606}      vs. 0.394 & $8.7 \times 10^{-8}$ & $l = 48$ (m2m) \tabularnewline
            $l = 48$ (m2m) &    0.523       vs. 0.477 & $2.6 \times 10^{-1}$ & $l = 512$ (m2m) \tabularnewline
            $l = 48$ (f2f) &    \textbf{0.581}       vs. 0.419 & $5.5 \times 10^{-5}$ & $l = 48$ (f2f)\tabularnewline
            $l = 48$ (f2f) &    0.513       vs. 0.487 & $5.1 \times 10^{-1}$ & $l = 512$ (f2f)\tabularnewline
            \hline \hline
        \end{tabular}
    }
\end{table}

\begin{table}[tb]\label{tab:eval-48kHz}
    \centering
    \caption{Preference scores with proposed lifter-training and conventional methods in full-band case \textbf{(48~kHz)}}
    \vspace{1mm}
    \subtable[\textbf{Speaker similarity}]{
        \footnotesize
        \begin{tabular}{r|cl|l}
            \hline \hline
            Proposed & Score & $p$-value & Conventional \tabularnewline
            \hline
            $l = 192$ (m2m) & 0.431 vs. \textbf{0.569}& $4.9 \times 10^{-4}$ & $l = 2048$ (m2m) \tabularnewline
            $l = 192$ (f2f) & 0.519 vs.   0.481       & $3.4 \times 10^{-1}$ & $l = 2048$ (f2f) \tabularnewline
            $l = 224$ (m2m) & 0.474 vs.   0.526       & $2.0 \times 10^{-1}$ & $l = 2048$ (m2m) \tabularnewline
            $l = 224$ (f2f) & 0.519 vs.   0.481       & $3.4 \times 10^{-1}$ & $l = 2048$ (f2f) \tabularnewline
            \hline \hline
        \end{tabular}
    }
    \subtable[\textbf{Speech quality}]{
        \vspace{-3mm}
        \footnotesize
        \begin{tabular}{r|cl|l}
            \hline \hline
            Proposed & Score & $p$-value & Conventional \tabularnewline
            \hline
            $l = 192$ (m2m) & 0.529 vs.     0.471       & $2.3 \times 10^{-1}$ & $l = 2048$ (m2m) \tabularnewline
            $l = 192$ (f2f) & 0.447 vs. \textbf{0.553}  & $8.9 \times 10^{-3}$ & $l = 2048$ (f2f) \tabularnewline
            $l = 224$ (m2m) & 0.513 vs.     0.487       & $5.2 \times 10^{-1}$ & $l = 2048$ (m2m) \tabularnewline
            $l = 224$ (f2f) & 0.517 vs.     0.483       & $4.2 \times 10^{-1}$ & $l = 2048$ (f2f) \tabularnewline
            \hline \hline
        \end{tabular}
    }
\end{table}
\begin{table}[tb]\label{tab:eval-subband}
    \centering
    \setlength{\tabcolsep}{1mm} 
    \caption{Preference scores with proposed sub-band processing and conventional methods in full-band case \textbf{(48~kHz)}}
    \vspace{1mm}
    \subtable[\textbf{Speaker similarity}]{
        \footnotesize
        \begin{tabular}{c|cl|c}
            \hline \hline
            Proposed (sub-band) & Score & $p$-value & Conventional \tabularnewline
            \hline
             m2m & 0.519 vs. 0.481 & $3.4 \times 10^{-1}$ & m2m \tabularnewline
             f2f & {\bf 0.603} vs. 0.397 & $5.0 \times 10^{-7}$ & f2f \tabularnewline
            \hline \hline
        \end{tabular}
    }
    \subtable[\textbf{Speech quality}]{
        \vspace{-3mm}
        \footnotesize
        \begin{tabular}{c|cl|c}
            \hline \hline
            Proposed (sub-band) & Score & $p$-value & Conventional \tabularnewline
            \hline
            m2m & {\bf 0.721} vs. 0.279 & $ < 10^{-10}$ & m2m \tabularnewline
            f2f & {\bf 0.700} vs. 0.300  & $ < 10^{-10}$ & f2f \tabularnewline
            \hline \hline
        \end{tabular}
    }
\end{table}

\vspace{-1mm}
\subsection{Subjective evaluations}
\vspace{-1mm}
\subsubsection{The evaluation of lifter training}\label{lifter_training}
\vspace{-1mm}
To investigate the effectiveness of the proposed methods, we conducted a series of preference AB tests on speech quality and XAB tests on speaker similarity of converted speech.  Thirty listeners participated in each of the evaluations through our crowd-sourced evaluation systems, and each listener evaluated ten speech samples. The target speaker's natural speech was used as the reference X in the preference XAB tests.
%We evaluated our two proposed methods. First, we compared the case where the filter was truncated using the proposed lifter-training method and the case using the conventional one without truncation. Table.~\ref{tab:eval-16kHz} shows the result in the narrow-band (16~kHz) VC. This figure indicates that proposed lifter-training method with the tap length 32 and 48 shows the same or higher quality (speaker similarity and speech quality) compared with the conventional method without truncation. This result demonstrate that the proposed lifter-training method can reduce the tap length to $1/16$ without degrading the speech quality in the narrow-band (16~kHz) case. The result in the full-band case is shown in Table.~\ref{tab:eval-48kHz}. This figure indicates that the proposed method ($l=224$) shows the same quality as the conventional one without truncation, and the proposed one ($l=192$) shows the lower quality. We can see that the proposed method can reduce the tap length significantly in the full-band case, though not as much as the narrow-band case.

We compared several settings of the conventional and proposed lifter-training methods. Table~1 lists the results for the narrow-band (16~kHz) case. Compared to the truncated conventional method (``Conventional ($l = 32, 48$)''), we can see that the proposed lifter-training method significantly outperformed the conventional one in terms of speaker similarity and speech quality. Also, compared to the non-truncated conventional method (``Conventional ($l =512$)''), the proposed lifter-training method (``Proposed ($l = 32, 48$)'') had the same or higher quality. These results indicate that the proposed lifter-training method can reduce the tap length to $1/16$ without degrading converted-speech quality whereas the truncated conventional method significantly degrades converted-speech quality. The same tendency can be seen in the full-band (48~kHz) case, as shown in Table~2. The proposed method with $l = 224$ had the same converted-speech quality as the non-truncated conventional method, but the proposed lifter-training method with $l = 192$ degraded speaker similarity and speech quality. Therefore, the proposed lifter-training method can significantly reduce the tap length in the full-band case, though not as much as the narrow-band case. 

\vspace{-1mm}
\subsubsection{The evaluation of sub-band processing}
\vspace{-1mm}
We also compared our proposed sub-band-processing method with the conventional method. The AB tests on speech quality and XAB tests on speaker similarity were conducted in the same manner as those mentioned in Section~\ref{lifter_training}. Table~\ref{tab:eval-subband} lists the results. Note that, lifter training and filter truncation were not used for both conventional and proposed methods. The proposed sub-band-processing method achieved a higher score than the conventional method excluding the speaker similarity of the male-to-male VC, demonstrating the effectiveness of this method.

\vspace{-1mm}
\section{Conclusion}
\vspace{-1mm}
We presented a lifter-training and sub-band-processing methods for computationally efficient and high-quality voice conversion based on spectral differentials. The lifter was trained considering filter truncation. The sub-band-processing method efficiently converted the lower frequency band of a full-band voice. The experimental results indicate the superiority of our methods in terms of computational efficiency, converted-speech quality compared to the conventional method. 
For future work, we will implement real-time VC using the proposed methods and evaluate its effectiveness in converted-speech quality and latency.
%

% acknowledgement
\textbf{Acknowledgements:} 
Part of this work was supported by the MIC/SCOPE \#182103104.

% bibliography
%\vfill\pagebreak
%\clearpage
\bibliographystyle{IEEEbib}
\bibliography{tts}

\end{document}